\documentclass[twocolumn,showpacs,preprintnumbers,prb,fleqn,floatfix]{revtex4}

\usepackage{graphicx}
\usepackage{dcolumn}
\usepackage{bm}

\begin{document}
\title{{Metal-to-insulator crossover and pseudogap in single-layer
compound Bi$_{2+x}$Sr$_{2-x}$Cu$_{1+y}$O$_{6+\delta }$ single
crystals in high magnetic fields }}

\author{S. I. Vedeneev$^{1,2}$ and D. K. Maude$^{1}$}

\affiliation{ $^1$ Grenoble High Magnetic Field Laboratory,
Max-Planck-Institut f\"{u}r Festk\"{o}rperforschung and Centre
National de la Recherche Scientifique, B.P. 166, F-38042 Grenoble Cedex 9, France \\
$^2$ P.N. Lebedev Physical Institute, Russian Academy of Sciences,
119991 Moscow, Russia}

\date{\today }

\begin{abstract}
The in-plane $\rho _{ab}(H)$ and the out-of-plane $\rho _c(H)$
magneto-transport in magnetic fields up to 28~T has been
investigated in a series of high quality, single crystal,
hole-doped La-free Bi2201 cuprates for a wide doping range and
over a wide range of temperatures down to $40$ mK. With decreasing
hole concentration going from the overdoped (p=0.2) to the
underdoped (p=0.12) regimes, a crossover from a metallic to and
insulating behavior of $\rho _{ab}(T)$ is observed in the low
temperature normal state, resulting in a disorder induced metal
insulator transition. In the zero temperature limit, the normal
state ratio $\rho _c(H)/\rho _{ab}(H)$ of the heavily underdoped
samples in pure Bi2201 shows an anisotropic 3D behavior, in
striking contrast with that observed in La-doped Bi2201 and LSCO
systems. Our data strongly support that that the negative
out-of-plane magnetoresistance is largely governed by interlayer
conduction of quasiparticles in the superconducting state,
accompanied by a small contribution of normal state transport
associated with the field dependent pseudogap. Both in the optimal
and overdoped regimes, the semiconducting behavior of $\rho _c(H)$
persists even for magnetic fields above the pseudogap closing
field $H_{pg}$. The method suggested by Shibauchi \textit{et al.}
(Phys. Rev. Lett. \textbf{86}, 5763, (2001)) for evaluating
$H_{pg}$ is unsuccessful for both under- and overdoped Bi2201
samples. Our findings suggest that the normal state pseudogap is
not always a precursor of superconductivity.
\end{abstract}

 \maketitle

\section{Introduction}

The electrical transport, notably the interlayer transport, of the
layered high-$T_c$ superconductors (HTS) shows anomalous
properties related to the quasi two-dimensional structure which
have been studied very extensively in recent years. In the normal
state the interlayer conductivity gives information on the
quasiparticle properties \cite{Morozov00}. This behavior of the
quasiparticles is anomalous in the normal state of the HTS, which
is of importance for elucidating the yet to be understood
mechanism of superconductivity. The understanding of the
fundamental interlayer transport properties of HTS is a
challenging physical problem in its own right.

One of the unusual features of the normal-state properties is the
coexistence of a metallic-like temperature dependence of the
in-plane resistivity $\rho _{ab}$ and a semiconducting-like
behavior for the out-of-plane resistivity $\rho _c$ (see e.g.
Refs. [\onlinecite{Martin,Brinceno,Foro}]). The very different
behavior of the resistivities $\rho _{ab}$ and $\rho _c$ implies a
2D confinement and is \textit{a~priori} incompatible with a
Fermi-liquid behavior \cite{Ando96}. Over the last few years, many
theoretical and experimental investigations have been devoted to
the transport properties of HTS. In particular, in the temperature
region showing the semiconducting-like $c$-axis resistivity, most
compounds reveal a negative out-of-plane magnetoresistance: the
Bi$_2$Sr$_2$CaCu$_2$O$_{8+\delta }$ (Bi2212)
\cite{Nakao,Yan,Wahl,Heine,Ando99,Morozov00}, the
La$_{2-x}$Sr$_x$CuO$_4$ (LSCO) \cite{Kimura,Hussey}, and the
La-doped Bi$_2$Sr$_{2-x}$La$_x$CuO$ _{6+\delta }$ (BSLCO) system
\cite{Ando96,Yoshizaki}.

The observed semiconducting-like $\rho _c(T)$ and negative
out-of-plane magnetoresistance have been discussed in terms of
different models, such as $c$-axis tunneling with a strong
suppression by charge fluctuations excited in the process of
tunneling \cite{Leggett}, $c$-axis hopping with interplanar
scattering \cite{Hussey}, a reduction of the density of states due
to superconducting fluctuations \cite{Nakao,Wahl,Heine}, and a
pseudogap and/or spin gap opening in the density of states
\cite{Yan,Kimura}. This is another striking feature of HTS. Of
particular interest in the physics of carriers in strongly
correlated and disordered systems to which HTS belong is the
coexistence of superconductivity and localization. The latter
phenomenon is one further peculiarity of HTS. Disorder in a
metallic system can cause localization of the electronic states
and lead to a metal-insulator transition \cite{Anderson}.

The metal-insulator transition has been observed in the
superconducting systems LSCO \cite{Boebinger} and
Pr$_{2-x}$Ce$_x$CuO$_{4+\delta }$ \cite{Fournier} at optimal
doping and BSLCO well inside the underdoped regime \cite{Ono00}.
The insulating behavior in these systems is characterized by an
in-plane resistivity $\rho _{ab}(T)$ which increases as $\log
(1/T)$. These results demonstrate that the metal-insulator
crossover in cuprates should not be universally associated with
doping but rather with the observation of a unified $\log (1/T)$
temperature dependence of the resistivity suggesting a peculiar
charge localization in the above mentioned cuprates \cite{Ono00}.

It is difficult to obtain an overall picture of the
metal-insulator transition in cuprates because only three systems
have been studied, with strikingly different results obtained for
the metal-insulator crossover for BSLCO when compared to those for
LSCO and Pr$_{2-x}$Ce$_x$CuO$_{4+\delta }$. The anomalous
transport should be more noticeable in the vicinity of the
metal-insulator transition and in the $T\rightarrow 0$ limit,
suggesting the existence of a close link between charge transport
and strong electron correlation. However, up to now the behavior
of cuprates in the normal state in the $T\rightarrow 0$ limit
remains an open issue.

One of the unresolved, but all-important issues of high
temperature superconductivity, is the connection of normal state
correlations cited above, and referred to as a pseudogap, to the
origins of the high $T_c$ \cite{Ding96}. Many experiments (e.g.
nuclear magnetic resonance \cite{Berthier}, photoemission
\cite{Ding96}, tunneling \cite{Renner}) have provided evidence
that in the normal state of underdoped HTS, a pseudogap exists in
the electronic excitation spectra below a temperature $T^{*}>T_c$.
This leads to a semiconducting-like behavior of the $c$-axis
resistivity below $T^{*}$. Photoemission experiments (ARPES) have
seen d-wave symmetry in the pseudogap structure \cite{Ding96}. In
scanning tunneling measurements on Bi2212, Renner $et$ $al.$
\cite{Renner} have found this pseudogap to be present both in
underdoped and overdoped samples, and to scale with the
superconducting gap. Certain groups have proposed, that the
pseudogap in the normal state can be seen as a precursor for the
occurrence of superconductivity where the superconducting
phase-coherence is suppressed by thermal or quantum fluctuations,
e.g. Refs.[\onlinecite{Emery,Hotta,Maly}]. More recently, from
interlayer tunneling spectroscopy in the Bi2212 system, evidence
for a definite difference between the superconducting gap and the
pseudogap has been obtained \cite{Suzuki}. This result is further
reinforced by nuclear magnetic resonance measurements \cite{Zheng}
on the underdoped cuprate YBa$_2$Cu$_4$O$_8$ ($T_c=74$ K) which
showed that a magnetic field of 23 T, while reducing $T_c$ by
23\%, has no effect on the pseudogap, suggesting that it has a
distinct origin from that of the superconductivity.

In the case of a non-superconducting origin, a pseudogap can be
formed in the spin-part of the excitation spectrum in the context
of spin charge separation. In studies of the magnetic field
dependence of the spin gap in the near optimally doped
YBa$_2$Cu$_3$O$_7$ in the normal state \cite{Gorny,Mitrovic}
probed using the spin lattice relaxation rate, contradictory
results were obtained. On the one hand, in an intensive study of
the anisotropic transport on the Bi2212 system \cite{Wantanabe97}
the authors found that the onset of semiconducting-like $\rho
_c(T)$ does not coincide with the opening of the spin gap seen in
the in-plane resistivity $\rho _{ab}(T)$. The pseudogap opening
temperature, on the other hand, coincides with the onset of the
semiconducting-like behavior observed in $\rho _c(T)$ on the
YBa$_2$Cu$_3$O$_7$ system. Since the normal-state properties in
the high-$T_c$ superconductors are known to depend strongly on the
carrier concentration, the reported transport and magnetotransport
data in the normal state cannot be easily categorized to form a
common picture. There is currently no consensus concerning at what
temperature the pseudogap opens \cite{Batlogg}. An experimental
investigation of the possible correlation between the pseudogap
and the out-of-plane magnetoresistance in layered HTS at high
magnetic fields is therefore of crucial importance.

In previous measurements \cite{Vedeneev00a} we have studied the
$c$-axis magnetoresistance in the La-free
Bi$_{2+x}$Sr$_{2-x}$Cu$_{1+y}$O$_{6+\delta }$ (Bi2201) single
crystals with $T_c=9.5$ K under magnetic fields up to 28 T and
over a temperature range $6-100$ K. The observed isotropic
behavior of the normal-state magnetoresistance with respect to the
orientation of the magnetic field (perpendicular and parallel to
the CuO$_2$ planes) shows that only the effect of the magnetic
field on the spins (Zeeman effect) is important in the normal
state. Such a result makes it difficult to explain the negative
magnetoresistance with models based on superconductivity involving
superconducting fluctuations or a pseudogap as a precursor of
complete superconductivity. Shibauchi \textit{et al.}
\cite{Shibauchi01} have reported $c$-axis resistivity measurements
in fields up to $60$ T in underdoped and overdoped Bi2212
crystals, from which they made a first evaluation of the pseudogap
closing field $H_{pg}$. These results again indicate the
predominant role of spins over orbital effects in the formation of
the pseudogap. However, because of the high $T_c=67-78$ K and very
high upper critical field, $H_{c2} $, for Bi2212 crystals the
available $60$ T field was insufficient to suppress
superconductivity at low temperatures and to evaluate $H_{pg}$,
the authors \cite{Shibauchi01} were forced to extrapolate their
data. Direct measurements of $H_{pg}$ were performed only at
$T>95$ K. So far as little is known about the effect of magnetic
field, the $H$ dependence of the pseudogap in HTS remains highly
controversial.

In this paper we present, to our knowledge, the first measured
temperature dependence for both the in-plane $\rho _{ab}$ and the
out-of-plane $\rho _c$ resistivities and magnetoresistivities
$\rho _{ab}(H)$ and $\rho _c(H)$ in hole-doped La-free Bi2201
cuprate at under, and optimal doping concentrations, and over a
wide range of temperature down to $40$ mK. Due to the lack of a
sufficient amount of Bi2201 single crystals and especially
crystals with different doping levels, the transport properties of
this system have not previously been investigated in detail. Owing
to the low critical temperature of Bi2201, $25$ T magnetic fields
are sufficient to suppress superconductivity in these samples in
the $T\rightarrow 0$ limit, even at optimal doping
\cite{Vedeneev99}. We have suppressed superconductivity in single
crystals using a 28~T resistive magnet at the Grenoble High
Magnetic Field Laboratory, in order to measure the in-plane
$R_{ab}$ and the out-of-plane $R_c$ resistances in the normal
state in magnetic fields applied perpendicular and parallel to the
$ab$-plane.

\section{Experiment}

\begin{table*}
\caption{\label{tab:Table1}Summary of the properties of the
investigated single crystals determined as described in the text:
The carrier concentration per Cu atom ($p$), actual cationic
compositions (Bi:Sr:Cu), ratios Bi/Sr, critical temperature
($T_c$), lattice parameter ($c$), disorder parameter ($k_Fl$),
pseudo gap closing field ($H_{pg}$), and the functional form of
the magnetic field dependence of $\rho_c(H)$.}
\begin{ruledtabular}
\begin{tabular}{cccccccc}
$p$  &  Bi:Sr:Cu  &  Bi/Sr  &  $T_c$ (K)  &  $c$ ($\AA$) & $k_Fl$ & $H_{pg}$ (T) & Functional form of $\rho_c(H)$\\
\hline
0.12  & 2.66:1.33:0.85  &  2.0  &  2.3  &  24.57 & 0.6 & $\geq30$ & $\rho _c(H) \simeq \rho_{c0}+a_{1}H$\\
0.13  & 2.62:1.38:0.87  & 1.9   &  3    &  24.575 & 7 & $\geq30$ & $\rho _c(H) = \rho_{c0}+a_{2}H+b_{2}H^2$\\
0.16  & 2.39:1.61:1.02  & 1.48  &  9    &  24.59 & 20 & $\simeq 21$ & $\rho _c(H,T)=\rho_{c0}+a_{3}\exp (-H/b_{3}T)$\\
0.17\footnote{Complete $\rho_{ab}(H)$ and $\rho_c(H)$ data is unavailable for this sample so that we are unable to estimate all parameters.}  & 2.31:1.69:1.12  & 1.37  &  9.6  &  24.61 & - & - & -\\
0.2   & 2.10:1.90:1.14  & 1.1   &  6.7  &  24.63 & 49 & $\simeq 16$ & $\rho _c(H,T)=\rho_{c0}+a_{4}\exp (-H/b_{4}T)$\\
\end{tabular}
\end{ruledtabular}
\end{table*}

It is known, that the stoichiometric composition Bi2201 is an
insulating phase, and that single-phase superconducting crystals
can be obtained by replacing Sr with either Bi or La \cite{Maeda}.
In a compound, the optimal cation states for Sr, La and Bi, are
Sr$^{2+}$, La$^{3+}$ and Bi$^{3+}$, respectively. Therefore, the
substitution of trivalent La or Bi for divalent Sr in the BSLCO or
in the La-free Bi2201 samples reduces the hole concentration in
the CuO$_2$ planes. For the Bi2201 samples, Fleming \textit{et al.
}\cite{Fleming} and Harris $et~al.$ \cite{Harris} found that as
the Bi/Sr ratio increases, and one moves toward the bottom of the
phase diagram of the solid solution, the number of holes doped
into the system decreases, which thus pushes the system towards
the hole-underdoped regime. The lower $T_c$, together with the
larger residual resistivity of Bi2201 in comparison with BSLCO
(the maximum $T_c$ is 38 K \cite{Ono00}) apparently suggests that
the disorder due to (Sr,Bi) substitution is stronger in Bi2201
than the disorder due to (Sr,La) substitution \cite{Ono03}.

We were able to make high quality single-phase superconducting
Bi$_{2+x}$Sr$_{2-x}$ Cu$_{1+y}$O$_{6+\delta }$ single crystals in
the range of $ 0.1<x<0.7 $, provided that the Cu content was
slightly increased \cite{Gorina94,Martovitsky}. The investigated
Bi2201 single crystals were grown by a KCl-solution-melt free
growth method. A temperature gradient along the crucible results
in the formation of a large closed cavity inside the
solution-melt. In this case, the crystals are not in direct
contact with the solidified melt in the crucible, thereby avoiding
thermal stresses during cool down. The crystals were grown in the
temperature range $830 -  850~{}^{\circ}$C. The crystals had a
platelet-like shape and mirror-like surfaces. The several tens of
crystals grown in such a cavity, when characterized, are found to
have almost identical properties. The quality of the crystals was
systematically verified by measurements of the dc resistance, ac
susceptibility, X-ray diffraction and scanning electron
microscopy. To summarize the properties of the investigated
crystals, we have regrouped in Table \ref{tab:Table1} the data of
$p$ (carrier concentration per Cu atom), actual cationic
compositions, ratios Bi/Sr, $T_c$, and lattice parameters $c$.

The X-ray diffraction measurements were performed using a
double-axis diffractometer. A CuK$_{\alpha}$ radiation
monochromatized by a pyrolytic graphite crystal was employed. Both
$\theta$- and 2$\theta$-scans of the ($0 0 l 0$) sublattice
reflections and the ($0 0 l \pm 1$) satellite reflections were
used to assess structural perfection. These measurements were
carried out before and after low-temperature experiments in
magnetic fields. The half-width of the sublattice reflections in
the X-ray rocking curves for the optimally doped single crystals
consisting of two or three blocks did not exceed $0.3^{\circ}$,
whereas for the crystals consisting of one block only (with
dimensions of only $0.3\times 0.3$~mm$^2$) it was less than
$0.1^{\circ}$. This value is close to a resolution limit of a
diffractometer. Both the ($\theta-2\theta$)- and $\theta$- X-ray
diffraction profiles of the sublattice show no detectable
structural defects. Thus, it can be concluded that even the
sublattice contains no small-angle boundaries. For example, the
half-width  of both the main profile (0016) and the satellite
reflections  (00151), (00151)' in the X-ray rocking curves for the
heavily underdoped single crystal with p = 0.13 (with large the Bi
excess) was about $0.2^{\circ}$.

The composition of the crystals was studied using a Philips CM-30
electron microscopy with a Link analytical AN-95S energy
dispersion X-ray spectrometer. The actual cationic compositions of
each investigated crystal were measured at several different
places on the crystal and the scatter in the data was less than
7\%. Complementary measurements of our Bi2201 single crystal
composition performed at the Material Science Center, University
of Groningen (The Netherlands) have shown that our crystals are
slightly underdoped due to oxygen depletion.

The dimensions of the crystals were ($0.4-0.8)~mm \times
(0.5-1)~mm \times (3-10)~\mu$m. The $T_c$ value of the crystals
formed by our free growth method can be as high as $13$ K.
However, we have found that the highest quality superconducting
Bi2201 single crystals have a very narrow range of values of the
lattice parameters $a=5.360-5.385~\AA$ and $c=24.57 - 24.63~\AA$.
In this case the $T_c$ (midpoint) values of the crystals lie in
the region $3.5-9.5~K$ in agreement with previous studies
\cite{Sonder,Harris}. The transition width defined by the $10\%$
and $90\%$ points of the superconducting transition of crystals
ranged from $0.5$ to $1.7~K$.

It is known, that overdoping or underdoping of
Bi$_2$Sr$_2$CaCu$_2$O$_{8+\delta }$ can be achieved by cation
substitutions or by changes in the oxygen content
\cite{Villard,Ooi}. However, in the low-$T_c$ Bi-based phase
Bi2201, we have found that it is difficult to change the number of
holes, because it is difficult to change the oxygen content. We
have performed many attempts to change the $T_c$ of single
crystals, after changing the doping level, by means of an
annealing in oxygen or argon at different temperatures. However, a
careful characterization of the annealed samples revealed that
changes in $T_c$ greater than $\pm 1$ K, were always accompanied
by a severe degradation of the sample quality and the occurrence
of phase inhomogeneity in agreement with previous studies
\cite{Sonder}. Most likely this is due to the fact that our
crystals are close to the decomposition line. For this reason, in
the following measurements, we used only high quality
\textit{as-grown} single crystals. For the investigation, samples
with different $T_c$ values were obtained by growing crystals with
a different Bi content.

A four-probe contact configuration, with symmetrical positions of
the low-resistance contacts ($<1\Omega $) on both $ab$-surfaces of
the sample was used for the measurements of $R_{ab}$ and $R_c$
resistances. The temperature and magnetic field dependence of the
resistances $R_{ab}(T,H)$ and $R_c(T,H)$ were measured using a
lock-in amplifier driven at $\approx $10.7 Hz. The measured
resistances were then transformed to the respective resistivities
$\rho _{ab}$ and $\rho _c$ using the crystal dimensions and the
ratio of $R_2/R_1$ in the thin sample limit of the Montgomery
technique \cite{Logan}. For the low temperature magnetotransport
measurements, the crystals were placed directly inside the mixing
chamber of a Kelvinox top-loading dilution fridge and studied with
the magnetic field $H$ applied either parallel or perpendicular to
the $c$ -axis. A configuration with $\mathbf{H\perp J}$ and
$\mathbf{H\parallel J}$ for the in-plane transport current
$\mathbf{J}$ was used. For the out-of-plane transport current, the
magnetic field $H$ was applied both parallel to the $c$-axis and
parallel to the $ab$-plane in the longitudinal
($\mathbf{H\parallel c\parallel J}$) and transverse
($\mathbf{H\perp c\parallel J}$) configurations.

The carrier concentration per Cu atom, $p,$ in the Bi-based HTS
cannot be unambiguously determined because the Bi ion does not
have a fixed valency \cite{Idemoto}. However, Ando \textit{et al}.
\cite{Ando00} have shown that the normalized Hall coefficient
$R_HeN/V_0$ of various cuprates agree well in the temperature
range 150 - 300 K and the data of La$_{2-x}$Sr$_x$CuO$_4$, for
which $p$ is unambiguous, can be used to estimate the doping level
in other systems. Here $e$, $N$ and $V_0$ are the electronic
charge, the number of Cu atoms in the unit cell and the volume
associated with each Cu atom, respectively.

In order to estimate the carrier concentration in our samples,
following the method proposed by Ando \textit{et
al}.\cite{Ando00}, we have measured the Hall coefficient $R_H$ in
several crystals\cite{Vedeneev00b,Bel} and compared the magnitudes
of the normalized Hall coefficient \cite{Bel} with the values
reported for LSCO \cite{Hwang}. Subsequently, we estimated $p$ in
other samples using the empirical (nearly linear) relation between
the excess Bi, $x$, and $p$. In the inset of  Fig.~\ref{fig1}, we
show the values of $T_c$ (closed circles) plotted vs $ p $ for our
Bi2201 single crystals (the dashed line is shown a guide to the
eye). It was found that optimum doping occurs at $p\simeq 0.17$
below which $T_c(p)$ shows a rapid drop as for the BSLCO system
\cite{Ando00}. As can be seen in Fig.~\ref{fig1}, our samples are
basically in the optimally doped and underdoped side of the phase
diagram and the data show the well-known parabolic behavior.

\section{Metal-insulator crossover and absence of a $\mathbf{\log (1/T)}$
divergence in both $\rho _{ab}$ and $\rho _{c}$}
\subsection{In-plane resistivity $\rho _{ab}$}

\begin{figure}
\includegraphics[width=0.7\linewidth,angle=0,clip]{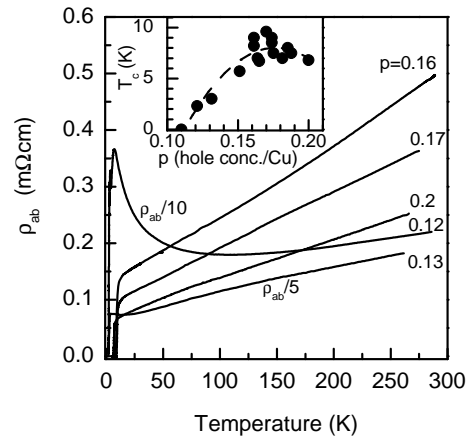}
\caption{\label{fig1}Temperature dependence of the in-plane
resistivity ($\rho_{ab}$) as function of temperature for Bi2201
samples with different hole concentrations. The inset shows the
critical temperature for superconductivity ($T_c$) determined from
$\rho_{ab}$ as a function of hole concentration.}
\end{figure}

In Fig.~\ref{fig1} (main panel) we show the temperature dependence
of the in-plane resistivity $\rho _{ab}$ for five single crystals
with $T_c=2.3$, $3$, $6.7$, $9.6$ and $9$ K (midpoint) at zero
magnetic field for $p$ values between $0.12$ and $0.2$. One can
see that as for other cuprates, the magnitude of $\rho _{ab}(T)$
increases with decreasing carrier concentration. The resistivity
curves give an almost linear temperature dependence for the
optimally doped sample, positive curvature for the overdoped
sample typical for other overdoped cuprates, and linear
temperature dependence for the underdoped samples with a
characteristic upturn at low temperatures (``semiconducting
behavior'').

\begin{figure}
\includegraphics[width=0.7\linewidth,angle=0,clip]{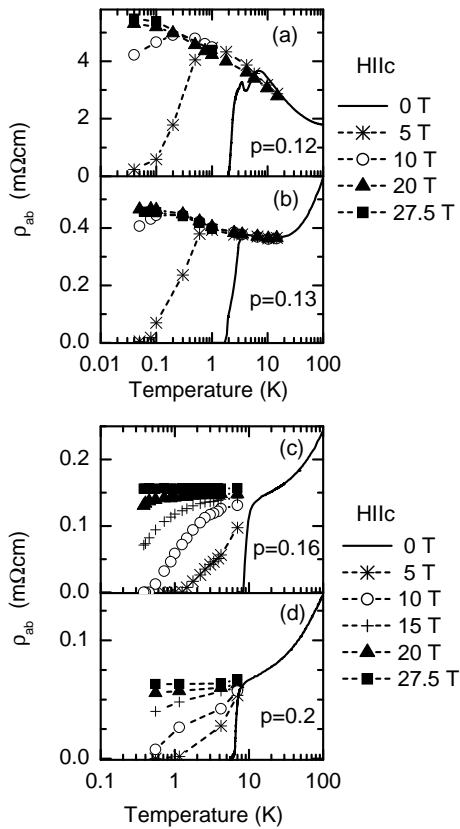}
\caption{\label{fig2} Semi-log plot of $\rho_{ab}$ versus
temperature for various magnetic fields applied along the c-axis
for four Bi2201 samples with different hole concentrations}
\end{figure}

Fig.~\ref{fig2} shows a semi-logarithmic plot of $\rho _{ab}(T)$
at various fixed magnetic fields for selected samples from
Fig.~\ref{fig1} in order to emphasize the low-temperature
behavior. Because the 20 and 27.5 T data are almost identical, we
believe that we are measuring the true normal-state resistivity at
our highest magnetic fields. $\rho _{ab}$ for two underdoped
samples, $p=0.12$ (a) and $0.13$ (b), goes through a minimum and
then at temperatures $T\approx 30$ K (a) and $T\approx 10$ K (b),
increases as $\log (1/T)$ as the temperature decreases, consistent
with the onset of localization \cite{Jing91}. This behavior is in
agreement with the results of Ono \textit{et al}. \cite{Ono00},
who found a logarithmic divergence of $\rho _{ab}(T)$ in
underdoped BSLCO and LSCO samples. The $\log (1/T)$ dependence of
$\rho _{ab}(T)$ reported by Ono \textit{et al}. \cite{Ono00}
extended over temperatures from $30$ to $0.3$ K without any sign
of saturation at low temperatures. However, as can be seen from
Fig.~\ref{fig2}(a) and (b), $\rho _{ab}$ in Bi2201 shows a
downward deviation from a $\log (1/T)$ dependence at ultra low
temperatures, $T=$ $0.04-0.2$ K, in very high fields. This
deviation cannot be related to the proximity of the
superconducting transition since the behavior of $\rho _{ab}(T)$
in magnetic fields of $20$ T and $27.5$ T in Fig. ~\ref{fig2}(a)
and (b) is identical. Moreover, the data at $27.5$ T in Fig.
~\ref{fig2}(b) actually lie below the $20$ T data. We interpret
the observed onset of the saturation of $\rho _{ab}$, as a
suppression of the localization by the magnetic field.

\begin{figure}
\includegraphics[width=0.7\linewidth,angle=0,clip]{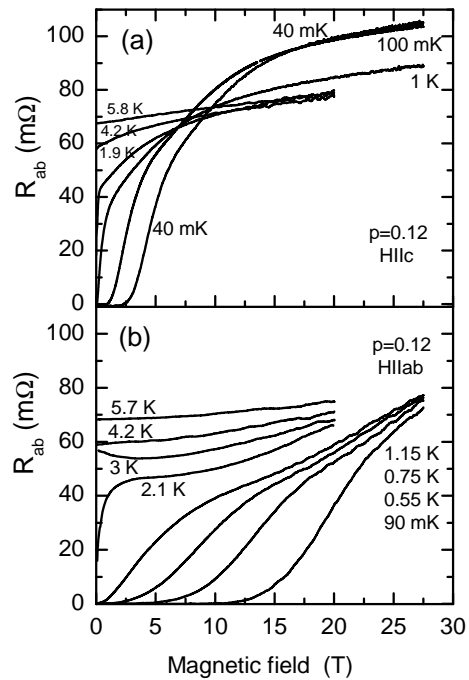}
\caption{\label{fig3} In-plane resistance as a function of
magnetic field applied along the c-axis (a) and in the ab plane
(b) measured at different temperatures for the underdoped Bi2201
sample with p=0.12.}
\end{figure}

One can see in Fig.~\ref{fig2}(a) that in the most underdoped
sample with $p=0.12$ at zero magnetic field there is the weak
upturn in the region $3-4$ K, which we believe is a consequence of
a competition between superconductivity and localization. To
illustrate this, we show in Fig.~\ref{fig3} the magnetic-field
dependence of $R_{ab}$ for the same sample with $p=0.12$ for
temperatures from $40$ mK to nearly $6$ K for magnetic fields
perpendicular (a) and parallel (b) to the $ab$-plane. The
considerable difference in the $R_{ab}(H)$ curves between the two
field orientations is a direct consequence of the anisotropy of
the upper critical field in Bi2201 due to a difference in the
orbital effect of the magnetic field on the one hand, and because
the effect of the magnetic field on the localization is weaker for
the parallel geometry on the other hand \cite{Lee}. As can be seen
from Fig.~\ref{fig3}(b), at $T=3$ K and $2.1$ K, a negative
magnetoresistance appears which results from the gradual
suppression of localization effects by the magnetic field. This
negative magnetoresistance also exerts some influence on the other
$R_{ab}(H)$ curves at lower $T$, that is to say, the localization
effects still persist. In the perpendicular geometry, the magnetic
field suppresses rapidly the superconductivity and the competition
between superconductivity and localization is not observed.
Although the localization also exerts some influence on the
$R_{ab}(H)$ curves in Fig.~\ref{fig3}(a) (curves have a pronounced
break-points in the derivative). Hence, we believe that the weak
upturn in the zero-magnetic field $\rho_{ab}$ in
Fig.~\ref{fig2}(a) is due to a competition between
superconductivity and localization. Nevertheless, a sample
inhomogeneity on atomic scale because of the heavy doping and
proximity of the isolating phase cannot be ruled out as a
possibility at this composition.

The negative magnetoresistance for the longitudinal geometry
itself presents an additional difficulty for standard interaction
theory. The same anomalous negative magnetoresistance for the
longitudinal geometry at low temperatures has been observed
previously in the nonsuperconducting Bi2201 single crystals by
Jing \textit{et al.}\cite{Jing91}. Since the authors \cite{Jing91}
considered this phenomenon in detail, we will not discuss this
topic further. However, it is important to note that in the second
most underdoped sample with $p=0.13$ the negative longitudinal
magnetoresistance is not observed in spite of the fact that the
$\rho _{ab}$ at $T < 10$ K in Fig.~\ref{fig2}(b) increases as
$\log (1/T)$ and the localization persists. The data in
Fig.~\ref{fig2} and Fig.~\ref{fig3} shows that the role of
disorder in the field-induced normal state of underdoped cuprates
remains an open question. Further experiments are needed to
reliably determine the low-temperature variation.

In contrast, $\rho_{ab}(T)$ for the slightly underdoped and
overdoped samples with $p=0.16$ Fig.~\ref{fig2}(c) and $0.2$
Fig.~\ref{fig2} (d) is constant below $5$ K and clearly shows a
metallic behavior in the normal state. This data is in full
agreement with the behavior of $\rho _{ab}(T)$ in BSLCO and LSCO
systems. Thus, it seems likely that the metal-insulator transition
in Bi2201 lies in the underdoped region ($p<0.16$) as for BSLCO.
The observed metallic behavior gradually changes to an insulating
behavior with decreasing carrier concentration.

In a 2D system the disorder parameter given by $k_Fl$, where $k_F$
is the Fermi wave vector and $l$ the elastic scattering length,
may serve as a measure of the disorder in the material
\cite{Fiory}. From the residual resistivity $\rho
_{ab}(T\rightarrow 0)$ in Fig.~\ref{fig2} and the lattice
parameter $c$ we determined the disorder parameter in the
$ab$-plane $(k_Fl)_{ab}\simeq 0.6,7,20,$ and $49$ for samples with
$p=0.12$, $0.13$, $0.16$, and $0.2$, respectively. For the samples
with $p\geq 0.16$, $(k_Fl)_{ab}>>1$ and a true metallic conduction
in the CuO$_2$ layers takes place, whereas the sample with p=0.12
clearly shows $\log (1/T)$ behavior starting from $T\simeq 30$ K
where the value of $\rho _{ab}$ is consistent with
$(k_Fl)_{ab}=1.3$ (it is important to note that the Mott limit
corresponds to $k_Fl=1$).

According to the optical data obtained by Tsvetkov \textit{et al}.
\cite{Tsvetkov} on our Bi2201 single crystals, the effective mass
in the $ab$-plane is $m^{\ast}=3m_{o}$ where $m_{o}$ is the
free-electron mass. Using this value of $m^{\ast }$ together with
the carrier density we can calculate $k_{F}$ and hence $l=60$ and
$145~\AA$ at 10 K for the samples with $p = 0.17$ and $0.2$,
respectively. This clearly indicates that the optimally doped and
underdoped Bi2201 crystals are clean superconductors. For these
calculations we have assumed a cylindrically shaped Fermi surface
with a highly anisotropic dispersion relation \cite{Kresin}.

The large increase of $\rho _{ab}$ is striking when compared with
the small change in $T_c$ when the hole doping $p$ is changed from
$0.13$ to $0.12$. This phenomena is not observed in the BSLCO
system, which supports the suggestion of Ono \textit{et al}.
\cite{Ono03}, that the disorder associated with (Sr,Bi)
substitution is more harmful to the electronic system than the
disorder due to (Sr,La) substitution. It is also possible that
this results from the proximity of the isolating phase near the
bottom of the phase diagram.

\subsection{Out-of-plane resistivity $\rho _{c}$}

\begin{figure}
\includegraphics[width=0.7\linewidth,angle=0,clip]{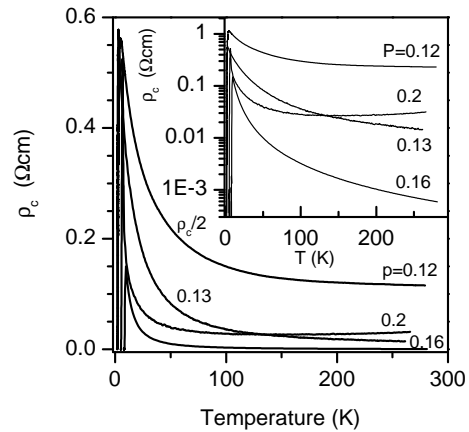}
\caption{\label{fig4} c-axis resistivity $\rho_{c}$ as a function
of temperature for Bi2201 samples with different hole
concentrations. The inset shows the same data plotted on a
semi-log scale.}
\end{figure}

Fig.~\ref{fig4} (main panel) shows the temperature dependence of
the out-of-plane resistivity $\rho _c$ at zero magnetic field for
four single crystals shown in Fig.~\ref{fig1}. The inset in
Fig.~\ref{fig4} plots $\rho _c(T)$ on a semi-logarithmic scale to
equally show the behavior of all the samples. As for the case of
$\rho _{ab}(T)$, with decreasing $p$, the overall magnitude of
$\rho _c$, increases as its ``semiconducting'' temperature
dependence becomes less marked. The exception is the overdoped
sample with $p=0.2$, for which the $\rho _c$ value is larger than
$\rho _c$ of the sample with $p=0.16$ and this sample already
shows a ``metallic'' temperature dependence of $\rho _c$ at high
temperatures. Such behavior at high temperatures is often observed
in overdoped cuprates. The larger value of $\rho _c$ in the
overdoped sample Bi2201 is likely due to an excess of Bi and
suggests a larger disorder in the electronic system compared to
the in plane disorder in the same sample probed by $\rho _{ab}$.
In all the underdoped crystals studied, we found that $\rho _c(T)$
at $H=0$~T varies as a power law $T^{-\alpha}$ over the
temperature range $T=3-300$~K with $\alpha =0.7-1.6$.

\begin{figure}
\includegraphics[width=0.7\linewidth,angle=0,clip]{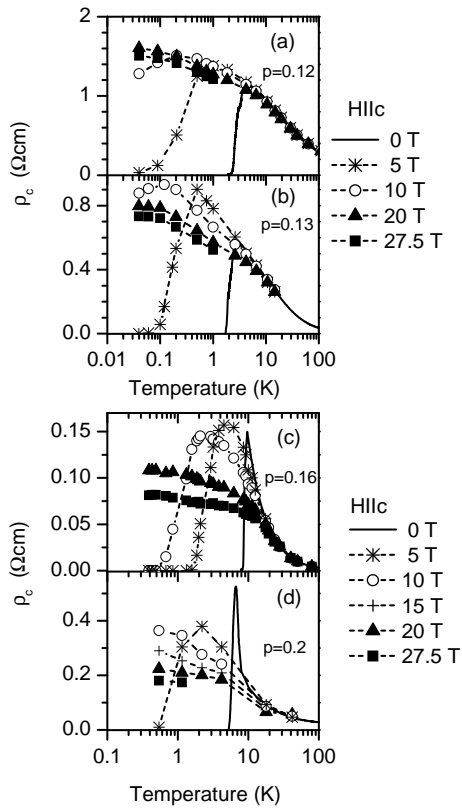}
\caption{\label{fig5} $\rho_{c}$ as a function of temperature and
measured at various magnetic fields applied parallel to the c-axis
for Bi2201 samples with different hole concentrations }
\end{figure}

A $\log T$ plot of $\rho _c$ at various fixed magnetic fields for
samples from Fig.~\ref{fig4} is shown in Fig.~\ref{fig5} in order
to emphasize again the low-temperature behavior. A strong
magnetic-field induced suppression of the low-temperature upturn
can be observed. In addition, $\rho _c(T)$ for the case of the
slightly underdoped or overdoped crystals shows a tendency to
saturate. One can see that the $\log (1/T)$ behavior of the $\rho
_c$ in the normal state gradually changes to a metallic-like
behavior with increasing carrier concentration. The onset of this
behavior in $\rho _c(T)$ moves to higher temperatures with
increasing carrier concentration. Our data in Fig.~\ref{fig5} are
in striking contrast to the behavior of $\rho _c(T)$ reported for
the underdoped LSCO samples \cite{Ando95} and the slightly
overdoped BSLCO single crystals \cite{Ando96}, which exhibited a
$\log (1/T)$ divergence in the normal state at $T\ll T_c$ (for
temperatures up to 0.66 K). The metallic-like temperature
dependence of the in-plane resistivity $\rho _{ab}$ and a
semiconducting-like behavior for the out-of-plane resistivity of
$\rho _c$ reported by Ando \textit{et al}. \cite{Ando96} suggested
that the $c$-axis transport is uncorrelated with the in-plane
transport. On the other hand, the same $\log (1/T)$ divergence of
$\rho _c(T)$ and $\rho _{ab}(T)$ in the underdoped LSCO samples
gave the authors of Ref.[ \onlinecite{Ando95}] additional evidence
against 2D localization. However, as is clear from
Fig.~\ref{fig5}, we do not have any evidence for a $\log (1/T)$
divergence at low temperatures in underdoped Bi2201 single
crystals, and the out-of-plane resistivity $\rho _c$ of the
slightly underdoped and overdoped Bi2201 single crystals below
$T_c$ in the highest applied fields shows almost no temperature
dependence. This implies that the carrier-transport mechanism in
the low-temperature limit, $T / T_c \to 0$, is the same for the
$ab$ and $c$ directions. We note that Morozov \textit{et al.}
\cite{Morozov00} have also observed a near saturation of $\rho _c$
for a Bi2212 crystal in the temperature region $22.5 - 30$~K at
55~T.

A parameter more often used than $\rho_c$ to characterize the
interlayer coupling is the anisotropy of the resistivity $\rho_c /
\rho_{ab}$. The largest anisotropy ratio found here is $\rho
_{c}/\rho _{ab}$ is $2.2\times 10^{4}$ just above $T_c$. We find
that the anisotropy ratio in zero-magnetic field for samples is
strongly temperature dependent except for the most underdoped
sample with $p=0.12$ for which $\rho _{c}/\rho _{ab}$ is
significantly less and depends only slightly on temperature,
probably due to the localization or enhanced disorder at this
doping level. Such behavior is in agreement for the most part with
the results of Wang \textit{et al.} \cite{Wang} and Ando
\textit{et al.} \cite{Ando96,Ono03} previously reported for BSLCO
samples and implies that at high temperatures the mechanisms
governing transport along and perpendicular to the CuO$_2$ plane
are different. However, the normal-state anisotropy ratio $\rho_c
/ \rho_{ab}$ at low temperatures in very high magnetic fields
becomes practically temperature independent for all samples. This
behavior is in distinct contrast to Ref.[\onlinecite{Ando96}]
where $\rho_c / \rho_{ab}$ of BSLCO crystals continued to increase
below $T_c$ providing evidence for the non-Fermi-liquid nature of
the system. On the other hand, this result is consistent with data
for the underdoped LSCO samples reported in Refs. [\onlinecite{
Boebinger,Ando95}]. The saturation of the ratio $\rho_c /
\rho_{ab}$ suggests that at low temperatures $\rho_{ab}$ and
$\rho_c $ in very high magnetic fields are related, which is
probably indicative of the anisotropic three-dimensional charge
transport in this region induced by the magnetic field. In view of
the remarkable difference between the temperature dependence of
$\rho_c / \rho_{ab}$ in Bi2201, BSLCO and LSCO, we do not want
discuss here this topic more fully.

\subsection{Pseudogap}

\begin{figure*}
\includegraphics[width=0.9\linewidth,angle=0,clip]{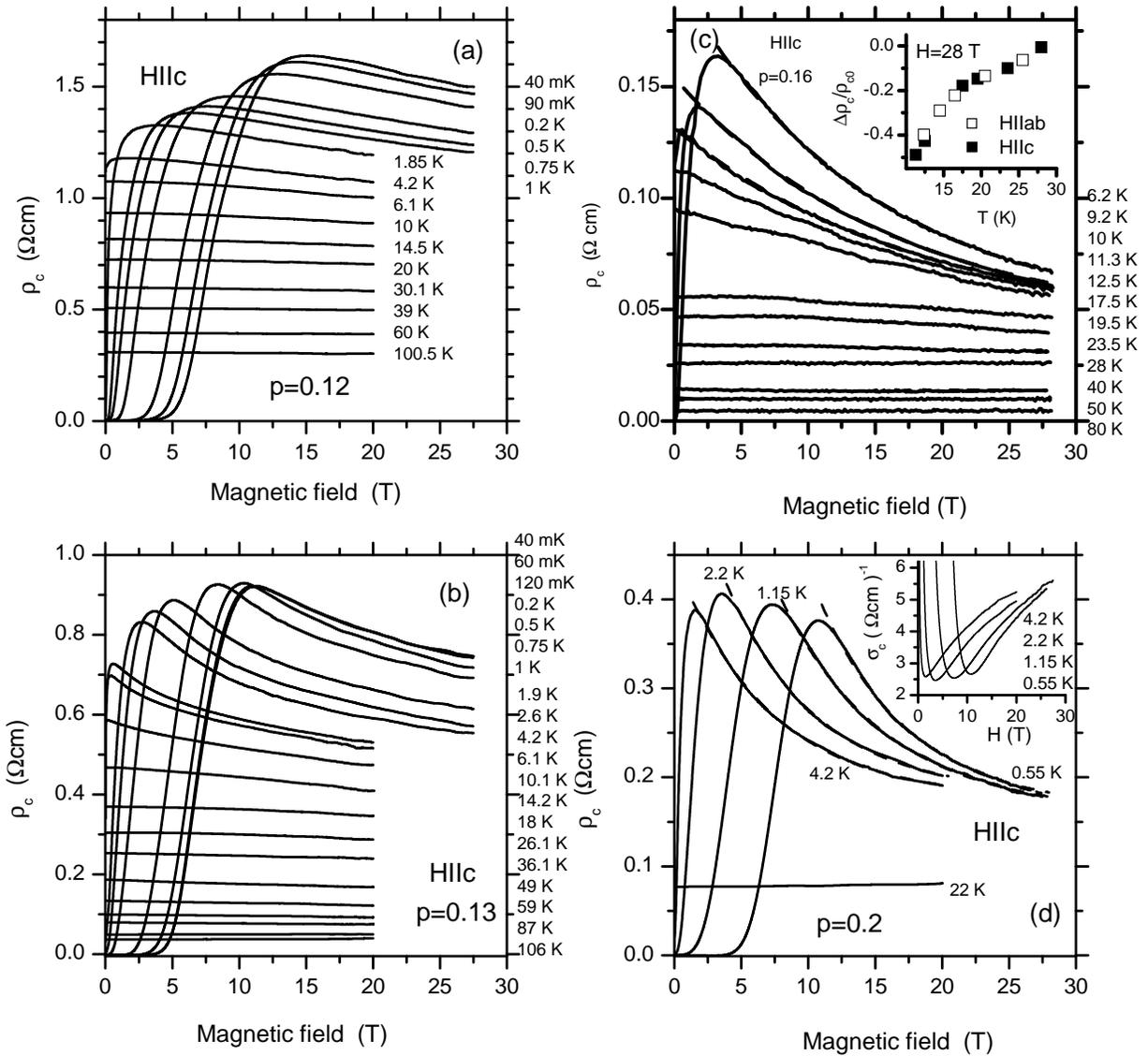}
\caption{\label{fig6} $\rho_{c}$ as a function of magnetic field
applied parallel to the c-axis for different temperatures for
Bi2201 samples with various hole concentrations. The inset in (c)
shows the relative variation $\Delta \rho_c / \rho_{c0} =
[\rho_c(H,T) -\rho_c(0,T) ] / \rho_c(0,T)$ at different
temperatures for both configurations at a magnetic field $H=28$~T.
The inset in (d) shows the c-axis conductivity for the same sample
($p=0.2$).}
\end{figure*}

According to Ref.[\onlinecite{ Morozov00}], the interlayer
transport results from a tunneling process and quasiparticle
tunneling dominates at higher fields. Since $\rho_c$ can give
information about the quasiparticle density of states in the
presence of a pseudogap, below we will discuss the $\rho_c$
magnetoresistivity at high fields in our samples. The suppression
of a semiconducting-like temperature dependence for $\rho_c(T)$
can be interpreted as the magnetic-field induced suppression of
the pseudogap, previously observed at temperatures above 5~K for
slightly underdoped Bi2201 crystals with $T_c=9.5$~K
\cite{Vedeneev00a} and in highly overdoped Bi2212 single crystals
\cite{Shibauchi03} at $T > 20$~K ($T_c\approx 60$~K).

In Fig.~\ref{fig6} we plot  $\rho _c(H)$ versus magnetic field for
four Bi2201 single crystals. For completeness in Fig.~\ref{fig6}
(c), we also display our data for the slightly underdoped
($p=0.16$) sample \cite{Vedeneev00a}. The inset in Fig.~\ref{fig6}
(c) shows the relative variation $\Delta \rho_c / \rho_{c0} =
[\rho_c(H,T) -\rho_c(0,T) ] / \rho_c(0,T)$ at different
temperatures for both configurations at a magnetic field $H=28$~T.
After the magnetic field induced onset suppression of
superconductivity all samples show a positive magnetoresistance at
low fields. The maximum in $\rho _c(H)$ observed at higher fields
is followed by a region of negative magnetoresistance.
Fig.~\ref{fig6} clearly shows the difference between the behavior
of $\rho _c(H)$ in the underdoped and overdoped crystals. At low
temperatures $\rho _c$ in the overdoped regime shows a much
stronger negative magnetoresistance compared to that observed in
the underdoped regime. Such a behavior of $\rho _c(H)$ has already
been intimated in the Bi2212 system \cite{Shibauchi01}. However,
such a large difference between the underdoped and overdoped
regimes in the slope of the negative magnetoresistance in
Fig.~\ref{fig6}, has not previously been observed. Furthermore, in
the heavily underdoped sample ($p=0.12$) after an increase of
$\rho _c$ at low fields due to the gradual suppression of
superconductivity, $\rho _c$ decreases almost linearly with
increasing magnetic field up to $\simeq 28$ T even at very low
temperatures in contrast to the power-law field dependence
previously reported in references
[\onlinecite{Morozov00,Shibauchi01}].

In Ref. [\onlinecite{Shibauchi01}] it was found that the field at
which the excess "semiconducting" resistivity $\Delta \rho _c(T)$
vanishes corresponds to the pseudogap closing field $H_{pg}$. A
fit to the power-law dependence of $\Delta \rho _c(H)$ for
magnetic fields above the maximum in $\rho _c(H)$ at different
temperatures allowed the authors of Ref.[\onlinecite{Shibauchi01}]
to find the field at which $\Delta \rho _c$ vanishes and evaluate
$H_{pg}(T)$ beyond the available 60 T. Based on this suggestion,
we tried to fit to a near linear field dependencies of $\Delta
\rho_c(H)$ in a log-log plot for $p=0.12$ and $p=0.13$ in order to
evaluate $H_{pg}$ at low temperatures in underdoped samples. This
evaluation gives exaggeratedly large values for $H_{pg}\approx
2000-3000$ T. In Ref.[\onlinecite{Shibauchi01}] it has also been
found that $H_{pg\text{ }}$and $T^{*}$ are related through the
Zeeman-like expression $g\mu _BH_{pg}=k_BT^{*}$, where $g=2$ is
the electronic $g$-factor, $\mu _B$ the Bohr magneton, and $k_B$
the Boltzmann constant. In our case such an analysis leads to
physically meaningless values $T^{*}=2700-4000$ K. Other
extrapolation polynomial fits gave the same physically meaningless
values of $H_{pg}$. These results probably indicate that the
method suggested in Ref. [\onlinecite{Shibauchi01}] for evaluating
$H_{pg}$ is unsuccessful in case of underdoped Bi2201 samples.

\begin{figure}
\includegraphics[width=0.7\linewidth,angle=0,clip]{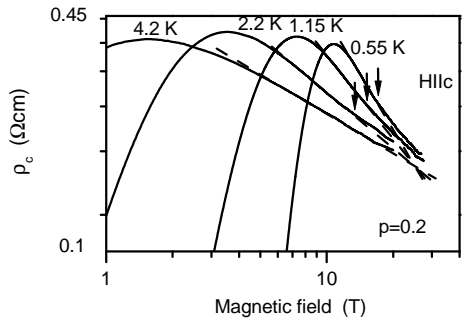}
\caption{\label{fig7} Log-log plot of $\rho_{c}$ versus magnetic
field for various temperatures for the overdoped Bi2201 samples.}
\end{figure}

We have tried to use such an extrapolation fit to our $\rho _c(H)$
data for overdoped Bi2201. Fig.~\ref{fig7} shows a log-log plot of
$\rho _c(H)$ at various fixed temperatures for the overdoped
sample with $p=0.2$. It can be seen that the dashed straight
lines, which are extensions of the linear dependencies, point to
the limiting value \cite{Shibauchi01} of $H_{pg}$, corresponding
to the intersection at $25$ T. If $H_{pg}\simeq 25$ T, then using
the Zeeman-like expression $T^{*}$ is found to be $\simeq 34$ K.
In the overdoped Bi2212 samples, the negative magnetoresistance
disappears at the same temperature at which the zero-field
$\rho_c(T)$ deviates from its characteristic linear (metallic)
high-temperature dependence \cite{Shibauchi01,Shibauchi03}. This
temperature in Ref. [\onlinecite{Watanabe00}] was identified as
the pseudogap closing (opening) temperature $T^{*}$. However, as
can be seen from Fig.\ref{fig4}, the zero-field $\rho _c(T)$ of
sample with $p=0.2$ deviates from a metallic linear (high)
temperature dependence at $T\simeq 140$~K, so that $H_{pg}$ should
be $\simeq 100$ T. The sample with $p=0.16$ does not show any
linear $T$ dependence (metallic state) up to $270$ K ( suggesting
that $H_{pg}$ should be $>200$ T). This result is clearly
inconsistent. Moreover, in Fig.~\ref{fig7} one can see that even
in case of overdoped Bi2201 samples there is only a finite range
of magnetic field for which the $\rho _c(H)$ data can be described
by a power-law \cite{Shibauchi01,Shibauchi03} dependence $H^\alpha
$ (dashed lines). This result indicates that the method suggested
in Ref. [\onlinecite{Shibauchi01}] for evaluating $H_{pg}$ is
unsuccessful in case of overdoped Bi2201 samples also.

Once the magnetic field at which the negative magnetoresistance
vanishes is identified with the pseudogap closing field $H_{pg}$,
our results clearly show that in the Bi2201 samples investigated
here, the pseudogap closing temperature $T^{*}$ does not agree
with the temperature at which the zero-field semiconducting-like
temperature dependence of $\rho_c$ changes into a metallic
dependence at higher temperatures as in the overdoped Bi2212
\cite{Shibauchi01}. Since the metallic-like linear temperature
dependence of the $\rho _c$ at $H=0$~T is a consequence of the
high doping of the samples which is inevitably accompanied by a
severe degradation of samples quality, we are unable to reach an
unambiguous conclusion concerning the relation of $H_{pg}$ with
the deviation from the linear temperature dependence of $\rho_c$.

However, on the other hand, in our slightly underdoped ($p=0.16$)
and overdoped ($p=0.2$) Bi2201 crystals the negative
magnetoresistance vanishes and the magnetoresistance changes sign
at $\approx 28$~K, the inset in Fig.~\ref{fig6}(c), and at
$\approx 22$~K, Fig.\ref{fig6}(d). Thus, $T^{*}$ should be close
to these temperatures. According to the the Zeeman-like expression
(Ref.[\onlinecite{Shibauchi01,Shibauchi03}]) the pseudogap closing
field scales with $T^{*}$ as $g\mu _BH_{pg}=k_BT^{*}$ which
implies that $H_{pg}$ should be $\simeq 21$ T and $\simeq 16$ T,
respectively.

In the slightly underdoped ($p=0.16$) and overdoped ($p=0.2$)
samples, the strong negative magnetoresistance rapidly weakens
[Fig.~\ref{fig6}(c) and (d)] and clearly shows a saturation at
high fields after more than a two-fold decrease. When the
temperature-dependent data in Fig.~\ref{fig5}(c) and (d) are
compared with Figs.~\ref{fig6}(c) and (d), it can be concluded
that the observed negative magnetoresistance corresponds to a
suppression of the semiconducting-like behavior in $\rho _c(T)$,
which can in turn be interpreted as the magnetic-field induced
suppression of the pseudogap. In previous measurements
\cite{Vedeneev99} we have shown that all $\rho _c(H)$ curves for
Bi2201 single crystal with $T_c\simeq 7$ K (overdoped) have a
pronounced break-point in the derivative well above the $\rho
_c(H)$ peak, which shifts to higher fields with decreasing
temperature and at $T\simeq T_c$ disappears. The field position of
these break-points in the derivative coincide with the
$H_{c2}^{*}$ values determined from the $\rho _{ab}(H)$ curves.
The values of $H$ at which the log-log plot of $\rho _c(H)$
deviates from a linear magnetic field dependence in
Fig.~\ref{fig7} (shown by arrows) are in close agreement with the
$H_{c2}^{*}$ values for a Bi2201 sample \cite{Vedeneev99} with
$T_c\simeq 7$ K. As has been shown in
Refs.[\onlinecite{Shibauchi01,Shibauchi03}], $H_{pg}$ in Bi2212
does not depend on temperature for $T<T_c$ and as $T\rightarrow 0$
$H_{pg}$ and the upper critical field, $H_{c2}$, coincide. This
suggests again that the intersection points of the dashed straight
lines in Fig.~\ref{fig7} is not $H_{pg}$ as observed in Bi2212. On
the other hand, if $H_{pg}$ is determined from the disappearance
of the negative magnetoresistance and, as pointed out above,
$H_{pg}\approx 21$ T ($p=0.16$) and $H_{pg}\approx 16$ T
($p=0.2$), so that $H_{pg}$ and $H_{c2}^{*}$ in Bi2201 are closely
linked as in Bi2212.

Yurgens \textit{et al.}\cite{Yurgens} measured the
intrinsic-tunneling spectra of a La-doped Bi2201 ($T_c$=32 K) at
$T$ = 4.5 - 300 K in order to determine the pseudogap phase
diagram. Their phase diagrams show that for samples with $p$=0.16
and under, the pseudogap closing temperature $T^{*}$ is over 300
K. These temperatures lie outside the range of our measurements.
While for the overdoped sample with $p$=0.2, the value of
$T^{*}$=22 K found in our work agrees well with the pseudogap
phase diagram of Yurgens \textit{et al.}\cite{Yurgens}.

The negative magnetoresistance observed in our experiments show a
characteristic exponential decrease with magnetic field.
Fig.~\ref{fig6}(c) and (d), show numerical fits (the dashed
curves) calculated using the functional form $\rho _c(H,T)=\rho
_{c0}+a\exp (-H/bT)$, where $a$ and $b$ are constants. Our data in
the slightly underdoped, optimally doped and overdoped regimes are
well described by such a functional form. The possibility to
describe $\rho _c(H)$ by an exponential expression in $H/T$
implies the magnetic-field couples to the pseudogap via the Zeeman
energy of the spin degrees of
freedom\cite{Vedeneev00a,Shibauchi01}.

In previous measurements of near optimally doped Bi2201 single
crystals\cite{Vedeneev00a}, we have found an isotropic behavior of
the normal-state magnetoresistance with respect to the orientation
of the magnetic field (perpendicular and parallel to the CuO$_2$
planes) which showed that only the effect of the magnetic field on
the spins (Zeeman effect) is important in the normal state. Here a
negative magnetoresistance is observed for both geometries
($\mathbf{H\parallel c\parallel J}$ and $\mathbf{H\perp c\parallel
J}$) in all investigated crystals. In contrast to the
magnetoresistance in the superconducting state, the normal-state
magnetoresistance of $\rho _c$ is independent of the field
orientation with respect to the current direction. For the
slightly underdoped sample ($p=0.16$), this behavior can be seen
in the inset in Fig.~\ref{fig6} (c).

We observed the same behavior for the heavily underdoped sample
($p=0.12$). The similarity in the normal-state data for the two
field orientations, probably excludes an explanation of the normal
state negative out-of-plane magnetoresistance in terms of
superconductivity.

It should especially be pointed out that in slightly underdoped,
optimally doped and overdoped samples after the field induced
suppression of the superconductivity and pseudogap for $\rho _c$
in high fields, the value of $\rho _c$ (after the saturation of
the magnetoresistance) remains much higher than the expected
un-gapped value. \emph{A semiconducting-like temperature
dependence of the out-of-plane resistivity $\rho _c$ is partly
conserved even after the suppression of the negative
magnetoresistance at $H > H_{pg}$}. It seems reasonable to
conclude that the semiconducting-like temperature dependence of
$\rho _c$ is controlled not only by the magnetic-field-sensitive
pseudogap.

\begin{figure}
\includegraphics[width=0.7\linewidth,angle=0,clip]{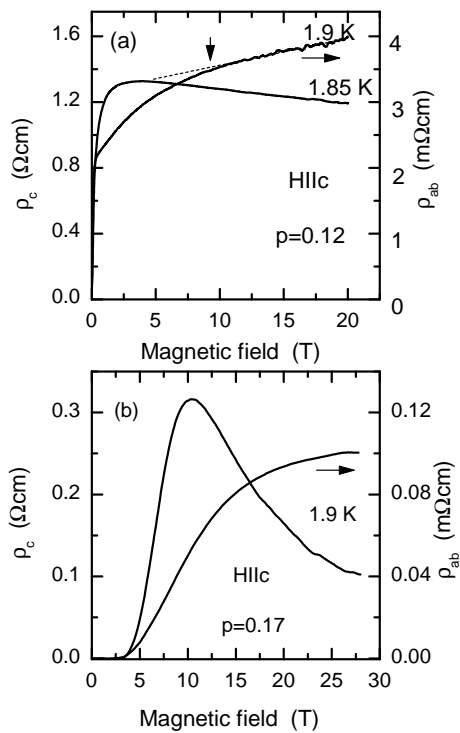}
\caption{\label{fig8} $\rho _c$ and $\rho _{ab}$ measured as a
function of magnetic field applied along the c-axis around T=1.9K
for the (a) underdoped and (b) optimally doped Bi2201 samples.}
\end{figure}

In previous measurements \cite{Vedeneev99} we have pointed out
that in overdoped samples ($T_c=7$ K) the maxima of $\rho_c(H)$
coincide with the field positions of $\rho_{ab}(H)
=0.4\rho_{ab}^n$ where $\rho _{ab}^n $ is the normal-state
resistivity. The observed maximum in $\rho_c(H)$ is a property of
the mixed state and results from a competition between the
``semiconducting'' behavior of $\rho_c$ and the superconducting
transition. In Fig.~\ref{fig8} we display the resistive
transitions of the heavily underdoped [(a), p=0.12] and optimally
doped [(b), p=0.17] samples in a magnetic field ${\bf H\Vert
c\Vert J}$ at temperatures $\approx 1.9$ K. It can be see that
after the maximum in $\rho_c(H)$, the $\rho_{ab}(H)$ curves still
show a strong positive magnetoresistance, clearly originating from
the superconducting state. In this range of magnetic fields a
superconducting gap still persists and part of the current along
the $c$-axis is a quasiparticle tunneling current. In the
underdoped sample with lower $T_c$ the superconducting gap is
small and closes rapidly when magnetic field is applied. The
resistive transition to the normal state is completed at $\approx
10$~T (the weak increase of the normal-state $\rho_{ab}$
resistivity is due to a magnetoresistance contribution at high
magnetic fields). The negative magnetoresistance at $H > 10$ T
displayed by $\rho_c(H)$ in Fig.~\ref{fig8}(a) is isotropic and is
due to the magnetic-field induced suppression of the pseudogap.
Since in the underdoped samples the magnitude of the pseudogap is
large, the effect of the magnetic field is small and the negative
magnetoresistance does not saturate in the available magnetic
field range [Fig.~\ref{fig6}(a),(b)]. In the optimally doped
[Fig.~\ref{fig8}(b)], slightly underdoped and overdoped
[Fig.~\ref{fig6}(c), (d)] samples because $H_{c2}$ is large, the
major contribution to the anisotropic negative magnetoresistance
in the $\rho_c(H)$ curves is due to the gradual decrease of the
superconducting gap (in Fig.~\ref{fig7}, the end of the
superconducting transition is indicated by arrows). Since in these
samples the magnitude of the pseudogap is small, the negative
magnetoresistance connected with the pseudogap rapidly saturates
following the superconducting transition [Fig.~\ref{fig6} (c),
(d)]. However, as previously pointed out the value of $\rho _c$
remains much higher than the expected un-gapped value. Recently it
has been shown that the negative magnetoresistance in the
superconducting state can also be described by the Zeeman-like
expression \cite{Pieri}. This explains why it is possible to
describe completely all curves $\rho _c(H)$ in Fig.~\ref{fig6}(c),
(d) using an expression which is exponential in $H/T$ in both the
superconducting and normal states.

\section{Conclusion}

We have presented the temperature dependence for both in-plane
$\rho _{ab}$ and out-of-plane $\rho _c$ resistivities and
magnetoresistivities $\rho _{ab}(H)$ and $\rho _c(H)$ in
hole-doped La-free Bi2201 cuprate for a wide doping range and over
a wide range of temperatures down to $40$ mK. We have shown that
the temperature and magnetic field dependence of the in-plane and
out-of-plane resistivities are determined by the localization, the
superconducting gap and the normal-state pseudogap. The data
suggest that the metal-to-insulator crossover in Bi2201 lies in
the underdoped region ($p<0.16$). The metallic behavior of $\rho
_{ab}(T)$ gradually changes to insulating behavior with decreasing
carrier concentration. We did not observe any evidence for a $\log
(1/T)$ divergence of $\rho _{ab}$ and $\rho _{c}$ at very low
temperatures in underdoped Bi2201 single crystals. The
out-of-plane resistivity $\rho _c$ of the slightly underdoped and
overdoped samples below $T_c$ in the highest applied fields shows
almost no temperature dependence. Our data strongly suggest that
the negative out-of-plane magnetoresistance appears to be governed
by different mechanisms; the main contribution comes from the
transition to the normal state which gives rise to a strong
magnetic field dependence, while the non-superconducting
pseudogap, shows a much weaker magnetic field dependence and
therefore only gives a small contribution to the negative
out-of-plane magnetoresistance. A semiconducting-like temperature
dependence of the out-of-plane resistivity $\rho _c$ is conserved
in part even after the suppression of the negative
magnetoresistance and at magnetic fields above the pseudogap
closing field $H_{pg}$. Our data support that the pseudogap does
not correlate with the existence of superconducting gap.

\begin{acknowledgments}
We thank V.P.Martovitskii for the careful X-ray studies of the
single crystals. This work has been partially supported by NATO
grant PST.CLG. 979896.
\end{acknowledgments}

\bibliography{resub_vedeneev}
\end{document}